\documentclass[11pt]{article}

\usepackage{amsmath,amssymb}
\usepackage{graphicx}
\usepackage{threeparttable,booktabs}
\usepackage{color}
\usepackage{colortbl}
\usepackage{xcolor}
\usepackage{multirow}
\usepackage{lscape}
\usepackage{float}
\usepackage{setspace}
\usepackage[figurename=Figure,tablename=Table]{caption}
\usepackage[a4paper, margin=1in]{geometry}
\usepackage{enumitem}
\usepackage[authoryear,round]{natbib}
\usepackage{hyperref}
\hypersetup{colorlinks=true,linkcolor=black,citecolor=black,urlcolor=black}

\newcommand{\indep}{\rotatebox[origin=c]{90}{$\models$}}

\DeclareMathOperator{\logit}{logit}

\title{Improving Survey Inference in Two-phase Designs Using Bayesian Machine Learning}

\author{%
Xinru Wang$^{1,2}$,
Anyu Zhu$^{1}$,
Lauren Kennedy$^{3}$,
Abigail Greenleaf$^{4}$,
Qixuan Chen$^{1,*}$\\[0.6em]
\parbox{\linewidth}{\footnotesize\sloppy\centering
$^{1}$Department of Biostatistics, Columbia University Mailman School of Public Health, New York, NY, USA\\
$^{2}$Centre for Quantitative Medicine, Duke-National University of Singapore Medical School, Singapore\\
$^{3}$School of Computer and Mathematical Sciences, University of Adelaide, South Australia, Australia\\
$^{4}$Department of Population and Family Health, Columbia University Mailman School of Public Health, New York, NY, USA\\[0.4em]
$^{*}$Corresponding author: \href{mailto:qc2138@cumc.columbia.edu}{qc2138@cumc.columbia.edu}
}}

\date{}

\begin{document}

\maketitle

\begin{abstract}
The two-phase sampling design is a cost-effective strategy widely used in public health research. Analyzing the Phase II sample often involves creating subsample-specific weights. However, these weights can be highly variable, leading to unstable weighted analyses. Alternatively, the rich data collected during the first phase can be leveraged to improve survey inference for the Phase II sample. In this paper, we propose a Bayesian tree-based multiple imputation (MI) approach for estimating population means using the Phase II sample, where the parent survey was conducted using a complex survey design. The design features of the parent survey, such as strata and clusters, are incorporated into the tree-based imputation models.  Through simulations, we demonstrate that the tree-based MI method outperforms traditional weighted estimators, yielding smaller bias, lower root mean squared error, and narrower 95\% confidence intervals, with coverage rates closer to the nominal level.  Furthermore, we show that Rubin’s variance estimation method provides valid statistical inference for population mean estimation in our setting. We illustrate the application of the proposed tree-based MI method using data from a cellphone survey on COVID-19 vaccination in Uganda, which represents a subcohort sample drawn from the 2020 Uganda Population-based HIV Impact Assessment Survey.
\end{abstract}


\section{\large Introduction}\label{sec:intro}
\vspace{0.1in}

\begin{spacing}{2}

Over the past century, survey sampling has been used to estimate population characteristics in a wide range of scientific and industrial disciplines such as healthcare, economic policy, and agriculture \citep{1977Foundations, hassan2020efficient}. These estimates provide useful information for decision-making and policy formulation \citep{lavange2001applying}. However, if the desired measures are difficult or expensive to obtain, it may not be feasible to collect these measures from all survey subjects due to practical considerations and budget constraints. A two-phase sampling design can be used to overcome this challenge. The first phase selects a sample from a finite population using probability sampling, during which easy-to-access and inexpensive measures are collected. The hard-to-obtain and expensive measures are then collected on a subsample selected from the Phase I sample \citep{hughes1996cluster}. Take the Mental Health Surveillance Study (MHSS) \citep{karg2014past} as an example. The National Survey on Drug Use and Health (NSDUH) \citep{cotto2010gender} is a nationwide study that aimed to provide up-to-date information on tobacco, alcohol, and drug use, mental health and other health-related issues in the United States. MHSS used data from clinical interviews administered to a subsample of respondents of the NSDUH to make inferences of the prevalence of serious mental illness among people aged 18 years and older. Given that it is impractical to conduct clinical interviews within the entire NSDUH sample with approximately 46,000 adult participants, a Phase II subsample was selected to collect the serious mental illness data in the MHSS. 

Statistical analysis of the Phase II sample could generate biased inferences of population quantities if the distribution of the Phase II sample differs from that of the population. A common approach to reducing bias in the Phase II sample is weighting, which accounts for subsample selection and nonresponse bias. This method creates a new weight variable by multiplying the weights for the Phase I sample with the subsample weight adjustments  \citep{kalton1986handling,kalton1987effects,chen2015}. However, weighting has several limitations in two-phase designs. First, weighting adjustments can lead to highly variable weights for the Phase II sample by compounding the already variable weights for the Phase I sample with additional adjustments. Second, the Phase I data used for weighting adjustments primarily focus on correcting distributional differences between the Phase I and Phase II samples, rather than emphasizing variables associated with the survey outcomes of interest. This may result in less efficient survey estimates.  Third, weighting adjustments might produce biased estimates if the subsample propensity model used to create the Phase II weights is misspecified.

An alternative approach is to first multiply impute the outcome variables measured only in the Phase II sample for the survey subjects who participate in the Phase I but not the Phase II study. After imputation, the population means are then estimated by applying the survey weights created for the Phase I sample to the multiply-imputed Phase I data. Different from weighting, the rich information collected in the Phase I sample is directly used for imputing the unknown outcomes collected in the second phase and can significantly reduce bias and improve efficiency in survey estimates if the data collected in the Phase I sample contains important predictors of the outcomes of interest. 

Well-chosen imputation models can generate efficient estimates of population quantities  \citep{kalton1986handling,  kalton1987effects, chen2015}. However, imputation methods may perform poorly if the imputation model is misspecified. The risk of model misspecification motivates the use of imputation models that are less sensitive to misspecification. Bayesian Additive Regression Trees (BART), which was first proposed by \citet{chipman2007bayesian}, is a sum-of-trees machine learning model that has gained significant popularity in recent years \citep{chipman2010bayesian}. BART uses the sum of multiple trees, developed through Bayesian back-fitting with Markov chain Monte Carlo (MCMC), to model the posterior distribution of the missing variables. Unlike traditional models, BART allows the inclusion of a large number of predictors without requiring analysts to select only a few. It is less sensitive to model misspecification and accommodates nonlinear effects and multi-way interactions between auxiliary variables and outcomes of interest. BART has been shown to outperform other machine learning methods, such as boosting, neural networks, and random forests, for prediction tasks \citep{chipman2007bayesian}. In addition to binary and continuous outcomes, BART has been extended to handle different types of outcomes, including count and categorical responses \citep{murray2021log}, semi-continuous outcomes \citep{linero2020semiparametric}, and survival outcomes \citep{sparapani2016nonparametric}.

BART and its extensions have been applied to a wide range of research areas \citep{tan2019bayesian},  including but not limited to
causal inference \citep{hill2011bayesian, leonti2010causal, hahn2020bayesian}, missing data \citep{xu2016sequential,kapelner2015prediction}, and survey inference \citep{liu2023inference, rafei2022robust}. 
Building on the strengths of BART, \citet{tan2019robust} proposed incorporating BART into doubly robust estimators for missing data and showed that using BART could reduce bias and RMSE when both the propensity and outcome models are misspecified. Additionally, they introduced BARTps, which incorporates the response propensity estimated via BART as an additional predictor in the outcome model, and demonstrated that BARTps performs well, particularly when both the propensity and mean models are complex. Similarly, \citet{liu2023inference} investigated the use of BART and soft BART as prediction models to estimate population means in survey inference when high-dimensional auxiliary information is available in both non-random samples and the target population. They demonstrated that this approach produces efficient and accurate estimates with 95\% probability interval coverage rates close to the nominal level. Consistent with the findings of \citet{tan2019robust}, they also showed that including the estimated propensity score as a predictor improves inference when the predictive model is potentially misspecified.

\citet{liu2023inference} focused on a simplified setting involving inference from one-phase samples obtained through single-stage sampling from a finite population. In this paper, we extend their framework to a more general two-phase design, where the Phase I sample is drawn from a finite population using a stratified multistage sampling strategy, and the Phase II sample is subsequently drawn from within the Phase I sample. We propose a tree-based multiple imputation (MI) approach, using BART-based models to impute values in the Phase I sample for variables observed only in Phase II. The imputed Phase I data, combined with Phase I sample weights, are then used to estimate the population mean. Furthermore, we show how to incorporate complex survey design features, such as strata and clusters, into the tree-based imputation models.

A key challenge in applying Liu's approach to our setting is its focus on single-phase samples, where posterior samples can be directly used for inference.  In our two-phase design, however, this approach does not adequately account for the effects of the complex sample design used in Phase I. To address this, we adopt Rubin's variance estimation method for multiple imputation, which enables robust inference for population-level quantities. We assess the performance of the proposed tree-based MI approach in comparison to alternative weighting methods through an extensive simulation study. Additionally, we illustrate the practical application of the BART-based MI method using subcohort cellphone survey data from Uganda. 

\section{\large Application: estimating COVID-19 vaccination coverage from a subcohort cellphone survey in Uganda}\label{sec:Motivating}
\vspace{0.1in}


To assess the status of the HIV epidemic in Uganda, the 2020 Uganda Population-based HIV Impact Assessment Survey (UPHIA2020) was conducted from February 2020 to March 2021. Using a two-stage cluster sample design, the survey included 313 enumeration areas and was nationally representative of adults aged 15 years and older \citep{UPHIA2020}. A total of 10,527 eligible households were surveyed, with 26,071 individuals completing individual interviews. Survey weights were developed to ensure the UPHIA2020 sample was representative of the target population. During the consenting process, all participants were asked if they agreed to be contacted for future research. 

In March 2022 a subcohort of the sample was invited to participate in a survey on COVID-19 vaccination coverage in Uganda. The sampling frame included households with at least one UPHIA2020 participant aged 18 or older who consented to future research, reported cellphone ownership, and spoke one of seven survey languages (Ateso, English, Luganda, Lugbara, Luo, Runyankole-Rukiga, or Runyoro-Rutoro). Unlike the initial survey, only adults aged 18 or older were eligible. Within each enumeration area, households were sampled proportional to the number of households in each enumeration area in the frame. Enumeration areas with fewer than five households were fully sampled, with all households included in the cellphone survey. One eligible household member who owned a cellphone was randomly selected as the ``primary respondent''. To address coverage bias, a ``secondary respondent'' was randomly chosen from non-phone owners in the same household. Consenting secondary respondents participated in the cellphone survey after the primary respondent completed the survey and passed the phone to them.  

Respondents in the cellphone survey were asked three questions about the main outcome of interest, COVID-19 vaccination status.  First they were asked, ``Have you ever tried to get a COVID-19 vaccine for yourself?'' If the respondent answered yes, they were asked, ``Have you received any dose of the COVID-19 vaccine?'' For those who answered yes again, additional details were collected, including the date of vaccination and whether they received a second dose. Respondents could indicate that a second dose was not required if they received a single-dose vaccine. Individuals who reported receiving one dose of a single-dose vaccine or two doses of any other vaccine were classified as fully vaccinated. From this subsample, we aim to estimate the prevalence of COVID-19 vaccination among the adult population in Uganda.

\section{\large Methods}
\label{sec:methods}
\subsection{\normalsize Notations and background}
\label{sec:methods1}
\vspace{0.05in}

Let $\mathcal{P}$ represent a finite population of size $N$ with strata $h=1,\dots, H$, each of which has clusters $j = 1, \dots, N_{h}$, where $N_h$ is the number of clusters for the $h$-th stratum. Let $y_{i}$ be the survey outcome of interest, $i=1,\dots,N$. The estimand of interest is the population mean $Q(y)=\sum_{i=1}^{N} y_i/N$. Let $\mathcal{C}$ denote the Phase I sample comprising of $n_c$ individuals selected from $\mathcal{P}$ using a stratified cluster sampling design; and $\mathbf{x}_i$ and $\mathbf{z}_i \in \{0,1\}$ are vectors of continuous and binary covariates for the $i$-th individual, which are collected at the individual level in the Phase I sample $\mathcal{C}$. 

To make the sample $\mathcal{C}$ representative of the population $\mathcal{P}$, a Phase I sample weight, $w_{ci}$, is assigned to the sample unit $i$ to adjust for the unequal probability of selection and nonresponse. When $y_i$ is fully observed for all the sample units in $\mathcal{C}$, the population mean can be estimated using
\begin{equation}
\label{equation_1}
\hat{Q}_{c}(y)=\frac{\sum_{i=1}^{n_c}w_{ci}\cdot y_i}{\sum_{i=1}^{n_c} w_{ci}},
\end{equation}
with variance estimated using the Taylor series linearization \citep{rust1996} or the resampling methods \citep{Rust1985, rust1996, Wolter2007}, such as jackknife repeated replication, balanced repeated replication, or bootstrap. 

In the two-phase designs considered in this paper, however, $y_i$ is only measured among $n_s$ individuals who also participate in the Phase II study denoted by $\mathcal{S}$. Under this situation, Formula~(\ref{equation_1}) cannot be applied directly. The conventional approach is to assign a subsample weight $w_{si}$ for the subsample in the second phase, which is often calculated as the product of the Phase I weight $w_{ci}$ and a subsample adjustment reflecting the selection probability and nonresponse for the subsample units. With the subsample weight $w_{si}$, the population mean using the subsample is estimated as
\begin{equation}
\label{equation_2}
\hat{Q}_{\mathcal{S}}(y)=\frac{\sum_{i=1}^{n_s} w_{si}\cdot y_i}{\sum_{i=1}^{n_s} w_{si}}.
\end{equation}

The selection probability for the Phase II sample can be easily calculated based on the sampling procedure, and the nonresponse adjustment can be obtained using either adjustment cell methods or propensity score adjustment methods \citep{chapman1986nonresponse, david1983nonrandom,chen2015}. The adjustment cell approach assigns sample units into distinct cells based on the discretized auxiliary information on $\mathbf{x}$ and binary $\mathbf{z}$. For all sample units within each cell, the nonresponse adjustment is applied as the inverse of estimated response rate \citep{chapman1986nonresponse}. When dealing with a large number of auxiliary variables, the chi-square automatic interaction detection (CHAID) \citep{kass1980exploratory} is commonly used. CHAID selects important auxiliary variables and forms adjustment cells through a merge-and-split process \citep{kass1980exploratory}. For each variable, a chi-square test is conducted to test the independence between two different categories of a variable. If the chi-square test fails to reject the null with a user-specified alpha-level, these categories are merged into a single category. These merging steps are repeated until all of the categories are significantly different for each predictor. When splitting, the predictor with the smallest Bonferroni-adjusted p-value is chosen as the first split, and the splitting continues until: a) no significant predictors; b) the child node will have insufficient individuals if more splits are made; c) the number of individuals in some nodes is below a pre-specified number \citep{chen2015}. 

Another method of nonresponse adjustment is the propensity score adjustment. Logistic regression is commonly used to estimate the response propensity conditional on the Phase I sample 
\citep{rizzo1996comparison}:
\begin{equation}
 \logit(Pr(r_{i}=1 | \mathbf{x}_i, \mathbf{z}_i))=\gamma_0 + \mathbf{\gamma}_1^{T} \mathbf{x_{i}} + \mathbf{\gamma}_2^{T} \mathbf{z_{i}},
 \label{eq:rp}
\end{equation} 
where $r_i=1$ if a unit is included in the Phase II sample and $r_i=0$ if a unit is included in the Phase I sample but not in the Phase II sample, and $\mathbf{\gamma}_1$ and $\mathbf{\gamma}_2$ are vectors of coefficients for $\mathbf{x}_i$ and $\mathbf{z}_i$, respectively. Screening of response-related variables is often conducted before building the propensity model \citep{rizzo1996comparison}. Unlike adjustment cell methods, propensity score adjustment allows for the inclusion of both discrete and continuous variables as predictors of response propensity. However, this method has some limitations. First, it can result in very small response propensities, leading to large weighting adjustments and unstable estimation of population quantities. Second, it relies heavily on the correct specification of the response propensity model. If the model is misspecified, the resulting subsample weights will be biased.

\subsection{\normalsize MI method using the Bayesian Additive Regression Trees}
\label{sec:methods2}
\vspace{0.05in}
In this paper, we propose a tree-based MI method. Rather than approaching the challenge as an adjustment from the subsample to the Phase I sample, we frame it as a missing data problem in the Phase I sample. We employ a tree-based imputation model to impute missing data in the Phase I sample and then use the multiply imputed Phase I data to infer population quantities. While weighting methods rely on auxiliary variables collected in the Phase I sample to adjust weights for the Phase II sample, imputation methods utilize this information to impute missing outcomes. This approach can yield more accurate and efficient inferences, particularly when the auxiliary variables are strong predictors of the outcomes of interest.

We start with a BART model for continuous $y_i$ measured only in the Phase II sample. Because Phase I sample is drawn from a finite population using a stratified multistage sampling design, in addition to auxiliary variables $\mathbf{x}_i$ and $\mathbf{z}_i$ collected in the Phase I sample, we also include in the BART model the design variables, $strata_i$ for strata, $cluster_i$ for cluster indicators, and $\mathbf{w}_i$ for survey weights -- including Phase I sample weights $w_{ci}$ and subsample selection and nonresponse weighting adjustments $a_{i}$. For continuous $y$, the imputation model can be written as 
\begin{equation}
\begin{aligned}
     y_{i}&=G(\mathbf{x}_i, \mathbf{z}_i, strata_i, cluster_i,\mathbf{w}_i)+\epsilon_i \\
     & =\sum_{b=1}^{B}g\left(\mathbf{x}_i, \mathbf{z}_i, strata_i, cluster_i,\mathbf{w}_i; T_b,\mathbf{\mu}_b\right)+\epsilon_i, \quad \epsilon_i\ \sim N(0,\sigma^2)
\label{eqn:model4}
\end{aligned}
\end{equation}
where $B$ is the number of trees, $T_b$ is the structure of the b-th binary tree, $\mathbf{\mu_b} = \{\mu_{b,1}, \dots, \mu_{b, n_b} \}$ are the parameters assigned to each terminal node, $n_b$ is the number of nodes for the b-th binary tree, and $g(\cdot)$ is the function that links $\mathbf{\mu_b}$ to ($\mathbf{x}_i$, $\mathbf{z}_i$, $strata_i$, $cluster_i$, $\mathbf{w}_i$). 

One challenge for BART is the prior specification. \citet{chipman2010bayesian} simplified the prior specification by assuming that the components of each tree are independent and not related to the error term $\epsilon$. In addition, the priors for $p(T_b)$, $p(\mu_b|T_b)$, and $p(\sigma^2)$ need to ensure that every tree is a weak learner and prevent the model from overfitting and non-convergence. The prior for $p(T_b)$ is related with: i) the probability of being nonterminal for a node at depth $\rho = 0,1,2,\dots$, which is specified by $\alpha(1+\rho)^{-\beta}$, where $\alpha \in (0,1)$, $\beta \in [0,\infty)$; ii) the probability of being selected as a splitting variable; iii) for the selected splitting variable, the probability of splitting rules. Further, $\mu_{b,l}, l = 1,\dots, n_b$ follows a normal distribution $p(\mu_{b,l}|T_b) = N(\mu_{\mu}, \sigma_{\mu}^2)$ and $\sigma^2$ follows an inverse chi-square distribution  $\sigma^2\sim \nu \lambda / \chi_{\nu}^{2}$. To facilitate the derivation of the posterior distribution, the outcome $y$ is transformed into $\tilde{y} = \frac{y - \frac{\min(y)+\max(y)}{2}}{\min(y)+\max(y)}$ allowing the hyperparameter $\mu_{\mu}$ to be set as $0$ and $\sigma_{\mu} = \frac{0.5}{k \sqrt{B}}$, where $k$ is a hyperparameter that needs to be pre-specified.  \citet{chipman2010bayesian} suggest default values for the hyperparameters of these prior distributions and the number of trees $B$, although cross-validation can be used to select the optimal hyperparameters for these priors.

To account for the correlation between units from the same cluster in the Phase I sampling, we also consider an extension of BART known as random intercept BART (rBART)  \citep{tan_flannagan_elliott_2018}. Different from model (\ref{eqn:model4}) where $cluster_i$ is included as dummy variables in the model, the cluster effect is modeled using a random intercept in the rBART model:
\begin{equation}
\begin{aligned}
    y_{ji}&=G(\mathbf{x}_{ji},\mathbf{z}_{ji},strata_{ji},\mathbf{w}_{ji})+\delta_j+\epsilon_{ji}\\
    & =\sum_{b=1}^{B}g(\mathbf{x}_{ji},\mathbf{z}_{ji},strata_{ji},\mathbf{w}_{ji};T_b,\mu_b)+\delta_j+\epsilon_{ji},& \epsilon_{ji} \sim N(0,\sigma^2),
\end{aligned}
\label{eqn:model5}
\end{equation}
where the subscript $j$ denotes the $j$-th cluster, and $\delta_j$ is the random intercept for the $j$-th cluster with $\delta_{j} \sim N(0,\tau^2)$ and $\delta_{j} \indep \epsilon_{ji}$. Each cluster is seen as a group with the same random intercept. For a binary outcome, BART and rBART can be easily extended by the probit model: 
\begin{equation}
\begin{aligned}
    Pr(y_{i}=1|\mathbf{x}_{i},\mathbf{z}_{i},strata_i,cluster_i,\mathbf{w}_i) &=\Phi(G(\mathbf{x}_{i},\mathbf{z}_{i},strata_i,cluster_i,\mathbf{w}_i)) \\ 
    & \text{~or~} \Phi(G(\mathbf{x}_{ji},\mathbf{z}_{ji},strata_{ji},\mathbf{w}_{ji})+\delta_j)
    \label{eqn:model6}
\end{aligned}
\end{equation}
where $\Phi$ is the cumulative density function of a standard normal distribution. 

The $\mathbf{w}_{ji}$ in the imputation model is composed of two parts, the Phase I weights $w_{ci}$ and the Phase II sample weighting adjustments $a_i$. The weighting adjustment $a_{i}$ can be estimated using a binary (r)BART, with predictors including $\mathbf{x}_i$, $\mathbf{z}_i$, $w_{ci}$, $strata_i$ and $cluster_i$. If there is external information about the population, calibration methods such as raking could also be applied \citep{Deming1940}. 

Each set of posterior draws from models (\ref{eqn:model4})-(\ref{eqn:model6}) generates one imputation for the missing $y$ values in the Phase I sample. Let $\hat{y}_i^{d}$ represent the imputed value of $y$ for individual $i$ in the $d$-th imputation of the Phase I sample, where $d=1,\dots, D$ and $D$ is the total number of imputations. The combined value of $y$ for individual $i$ in the $d$-th imputation is expressed as: $\tilde{y}_{i}^{(d)}=I(i\in \mathcal{S})y_i+I(i\in \mathcal{C},\ i\notin \mathcal{S}){\hat{y}}_{i}^{d}$, where $I(\cdot)$ is the indicator function, which equals 1 if the condition within $\cdot$ is true and 0 otherwise. Using the $d$-th imputation, the population mean is then estimated by:
\begin{equation}\label{eq:singleImp}
    \tilde{Q}(y)^{(d)}=\frac{\sum_{i=1}^{n_c}w_{ci}\cdot{\tilde{y}}_{i}^{(d)}}{\sum_{i=1}^{n_c}w_{ci}},
\end{equation} with the estimated variance $\widehat{Var}  \left (\tilde{Q}(y)^{(d)} \right )$ obtained by the Taylor series linearization method to account for the sampling design and nonresponse in the Phase I sample \citep{rust1996}.

The $D$ imputations based on (\ref{eq:singleImp}) are then combined using the Rubin's multiple imputation rules with the variance accounting for both between-imputation and within-imputation variances \citep{little2002}:
\begin{equation}
\hat{Q}_{MI}(y)=\frac{1}{D}\sum_{d=1}^{D}\tilde{Q}(y)^{(d)}, ~~~\widehat{Var} \left (\hat{Q}_{MI}(y) \right ) =\bar{W}_D+(1+\frac{1}{D})B_D,
    \end{equation}    
where $\bar{W}_D = \frac{1}{D}\sum_{d=1}^{D}\widehat{Var}  \left (\tilde{Q}(y)^{(d)} \right )$ and $B_D = \frac{1}{D-1}\sum_{d=1}^{D}(\tilde{Q}(y)^{(d)}-\hat{Q}_{MI}(y))^2$ are the estimated within and between imputation variance, respectively. The reference distribution for confidence interval estimates is a $t$ distribution,
\begin{equation}
    \frac{\hat{Q}_{MI}(y)-Q(y)}{\sqrt{\widehat{Var} \left ( \hat{Q}_{MI}(y) \right )}} \sim t_{\nu},
\end{equation}
with the degrees of freedom based on a Satterthwaite approximation,
\begin{equation}
    \nu = (D-1)(1+\frac{D}{D+1}\frac{\bar{W}_D}{B_D})^2.
\end{equation}

Following \citet{liu2023inference}, the effectiveness of the tree-based imputation approach in two-phase designs relies on the following three assumptions: (A1) Outcome-related auxiliary variables are observed in the Phase I sample; (A2) The subsampling mechanism is conditionally ignorable, meaning that the outcome variable $y$ is independent of inclusion in the Phase II sample,  given all auxiliary variables and design features; and (A3) The auxiliary variables in the Phase I and Phase II samples cover comparable ranges. Assumption A3 is violated if subsampling results in the Phase I sample containing a broader range of auxiliary variable values than the Phase II sample. For example, when individuals with certain predictor values are consistently excluded from the Phase II subsample.  

\section{\large Simulation}
\vspace{0.1in}

We conduct a simulation study to evaluate the performance of both weighting and tree-based MI methods in a two-phase survey design. Four weighting methods are compared, with weights estimated using logistic regression (LGM), CHAID, BART, and rBART; the corresponding estimators are denoted as WT-LGM, WT-CHAID, WT-BART, and WT-rBART. In addition, we consider two MI methods based on tree-based models: MI-BART and MI-rBART. 

To evaluate the effectiveness of our proposed approach, we introduce a benchmark estimator, which represents the estimate of the quantity of interest under the assumption that the outcome $y$ is observed for all subjects in the Phase I sample. While this benchmark provides a useful point of comparison, it is important to note that it still carries errors, as it reflects the Phase I sample estimate rather than the true population value.

Additionally, we conduct a sensitivity analysis to assess the impact of incorporating subsample weighting adjustments, $a_i$, as part of $\mathbf{w}_i$ in the imputation models for the MI-based estimators, as defined in equations (\ref{eqn:model4})-(\ref{eqn:model6}). Specifically, we compare the performance of the MI-BART and MI-rBART estimators with and without $a_i$ included as a covariate in the imputation model. Furthermore, we evaluate differences in performance when $a_i$ is derived from the inverse of the true response propensities versus the estimated response propensities.

We consider various simulation scenarios and conduct $500$ replicates of simulation for each scenario. Model performance is evaluated based on absolute bias and root mean squared error (RMSE):
\begin{equation*}
\label{equation_3}
 \textrm{Absolute bias} = \left |{\frac{\sum_{s=1}^{500}(\hat{Q}(y)^{(s)}-Q(y))}{500}}\right |,
\end{equation*}

\begin{equation*}
\label{equation_4}
\textrm{RMSE} =   \sqrt{\frac{\sum_{s=1}^{500}(\hat{Q}(y)^{(s)}-Q(y))^2}{500}},
\end{equation*}
where $\hat{Q}(y)^{(s)}$ is an estimate of $Q(y)$ in the $s$-th simulation replicate, $s=1, \dots, 500$. We also investigate the coverage rate and the average width of the 95\% confidence intervals (CIs). These metrics are magnified a hundred times for reading convenience.

\subsection{\normalsize Simulation design}
\vspace{0.05in}

This subsection gives detailed steps to generate the two-phase multistage complex survey data. 

\begin{enumerate}[label=\arabic*.]
\item Population:
We first generate the finite population $\mathcal{P}$ with $4$ strata ($H=4$), each of which has $N_h = 25, 20, 15, 10$ clusters respectively. The number of individuals in each cluster $j=1,\dots,N_h$, denoted as $N_{hj}$, follows an exponential distribution truncated between $100$ and $300$, with an expected value of $200$. The overall population size is $N \approx 13,000$. This dataset serves as the finite population data across all simulation replicates. For each individual in the population, we generate both continuous and binary variables. We refer to the $l_1$-th continuous variable as $x_{l_1}$, $l_1=1,\dots, L_1$, and the $l_2$-th binary variable as $z_{l_2}$, $l_2 = 1, \dots, L_2$, where $L_1$ and $L_2$ are the numbers of continuous and binary variables, respectively. The continuous variable $x_{l_1}$ follows a standard normal distribution $N(0,1)$. Binary variables $z_{l_2}$ follow a Bernoulli distribution with $Pr(z_{l_2}=1)$ generated from a uniform distribution $U\left(0.4,0.6\right)$. 

\item Outcome model: We specify the true outcome model for the $i$-th subject in the $j$-th cluster of the $h$-th stratum as
\begin{equation*}
\begin{aligned}
     y_{hji} &=  2.47 + q_{hj}-2 x_{1hji}+x^2_{2hji}
    +2z_{1hji}-z_{2hji}-2z_{3hji} +x_{1hji}z_{1hji} + \epsilon_{hji}, \quad \epsilon_{hji} \sim N(0,1)
\end{aligned}
\end{equation*}
where $q_{hj}$ is the cluster random intercept which follows a normal distribution $N(0, 1)$. The true population mean of the outcome is $Q(y)=3$.

\item Phase I sample: To select the Phase I sample, we draw $n_h = 10, 8, 6, 4$ clusters respectively from each stratum using simple random sampling.
All the units in the selected clusters are invited to participate in the Phase I study but not all respond. The response probability for the $i$-th subject in the $j$-th selected cluster of the $h$-th stratum is:
\begin{equation}
\pi_{hji, res, 1} = \logit^{-1}(-1+2z_{1hji}+2z_{2hji}-z_{3hji}).
\end{equation}
This sampling procedure results in a Phase I sample with approximately $3,000$ individuals. 

\item Phase II sample: To collect outcome data, we select a subset of the Phase I sample by simple random sampling with a selection probability of 0.5. We consider five scenarios for the response patterns in the Phase II sample. These scenarios are inspired by, and extend, the simulation design presented in \citet{liu2023inference}:

\begin{enumerate}

   \item[S1:] Low dimensional auxiliary variables ($L_1=2$, $L_2=3$) collected in the Phase I sample. The true response propensity model is
    \begin{equation*}
    \begin{aligned}
         \pi_{hji, res, 2} &= \logit^{-1}(1+2x_{1hji}+1.5x_{2hji}^2+2 z_{1hji}+z_{2hji} -2z_{3hji}-x_{1hji}z_{1hji}).
    \end{aligned}
    \end{equation*}
    Units in the lower tail of $x_{1}$ and $x_{2}$ have a lower probability of response in the Phase II sample. 
    \item[S2: ] High dimensional auxiliary variables ($L_1=10$, $L_2=10$) collected in the Phase I sample. The true response propensity model is the same as S1, with the only difference that there are noise variables $x_3, \dots, x_{10}$, $z_4, \dots, z_{10}$ that are not associated with the outcome $Y$ or the response propensity.
    \item[S3: ] High dimensional auxiliary variables ($L_1=10$, $L_2=10$) collected in the Phase I sample with the assumption A3 violated in the outcome-related variables. The true response propensity model is specified as 
    \begin{equation*}
    \begin{aligned}
        \pi_{hji, res, 2} &= \logit^{-1}(1+2x_{1hji}-1.5x_{2hji}^2+2 z_{1hji}+z_{2hji}
        -2z_{3hji}-x_{1hji}z_{1hji}).
    \end{aligned}
    \end{equation*}    
    The only difference between S3 and S2 lies in the sign of the coefficient for $x_{2}^2$, now set to negative values such that there are sparse data in the higher and lower tails of $x_{2}$.
    
    \item[S4: ] High dimensional auxiliary variables ($L_1=10$, $L_2=10$) collected in the Phase I sample with the assumption A3 violated in the variables that are not related to the outcome. The true response propensity model is specified as 
    \begin{equation*}
    \begin{aligned}
        \pi_{hji, res, 2} &= \logit^{-1}(1+2x_{1hji}-1.5x_{3hji}^2+2 z_{1hji}+z_{2hji}-2z_{3hji}-x_{1hji}z_{1hji}).
    \end{aligned}
    \end{equation*}    
    In this scenario, the higher and lower tails of $x_{3}$ are under-sampled, leading to different ranges of $x_{3}$ in the Phase II and Phase I samples. Notably, unlike in S3, $x_{3}$ is not related to $y$.

    \item[S5: ] High dimensional auxiliary variables ($L_1=10$, $L_2=10$) collected in the Phase I sample, but none of the variables in the response model is related to the outcome:
\begin{equation*}
    \begin{aligned}
        \pi_{hji, res, 2} &= \logit^{-1}(1+2x_{3hji}+1.5x_{4hji}^2+2 z_{4hji}+z_{5hji}-2z_{6hji}-x_{3hji}z_{4hji}).
    \end{aligned}
    \end{equation*}    

\end{enumerate}
\end{enumerate}

\begin{figure}[H]
\centerline{\includegraphics[width=0.8\linewidth]{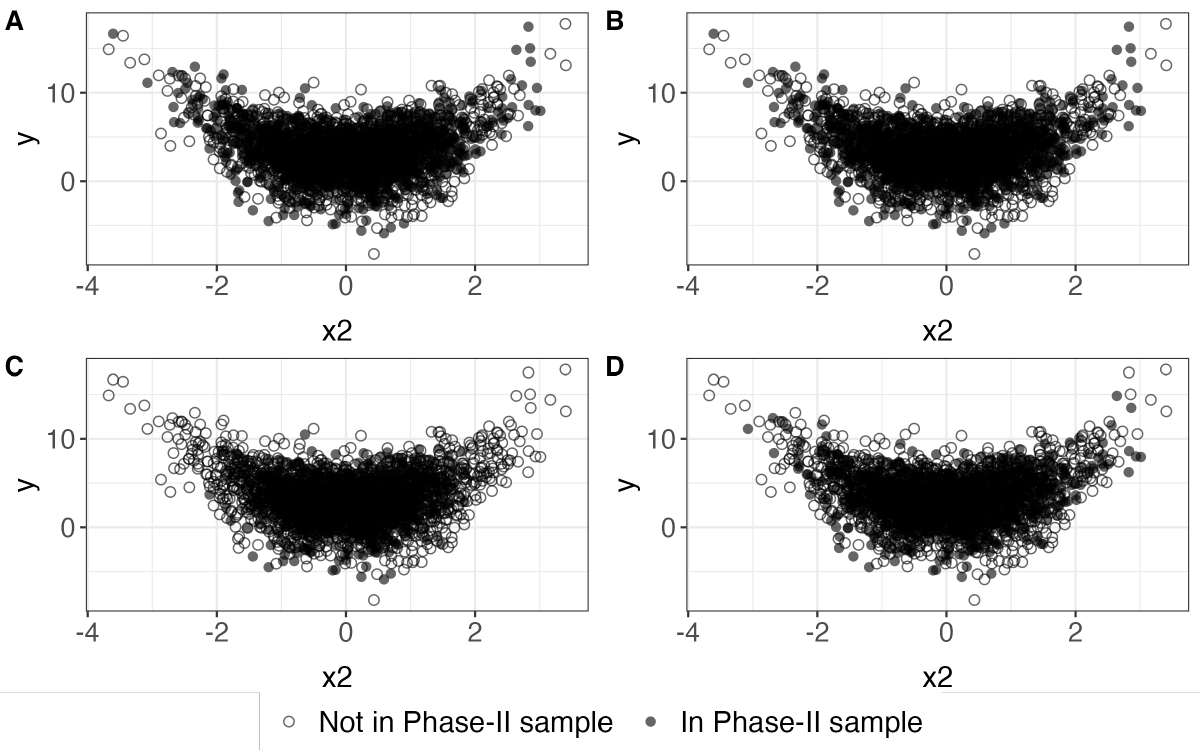}}
\caption{Scatter plots of outcome $y$ vs. continuous covariate $x_2$ across simulation scenarios.  Open circles represent subjects in the Phase I sample; dots indicate subjects included in the Phase II sample. Panels A-D correspond to simulation scenarios S1-S4, respectively. }
\label{scenarios}
\end{figure}

Figure~\ref{scenarios} shows scatter plots of Phase I and Phase II samples in one simulation replicate for scenarios S1-S4. The y-axis is the outcome $y$, and the x-axis is the continuous variable $x_2$ that is associated with $y$. The open circles represent the Phase I units; dots denote the Phase I units that are also in the Phase II sample. In scenarios S1, S2, and S4, the range of $x_2$ is similar in Phase I and Phase II samples; however, in scenario S3, the Phase I units in the higher and lower tails of $x_2$ are less likely to be included in Phase II. The scatter plot for scenario S5 is not shown here but it is similar to scenario S4 where the Phase II sample selection is not related to $x_2$.

Scenarios S2-S5 aim to evaluate the performance of the competing methods in real-world settings where there are a large number of auxiliary variables available in the Phase I sample. In these settings, researchers often lack clear insights into the truly important predictors for the response propensity or the desired outcome. To obtain WT-LGM and WT-CHAID estimates in these high dimensional settings, we first use Lasso \citep{tibshirani1996regression} to select key variables for the response propensity and then fit linear regression and CHAID models, respectively. Scenarios S3 and S4 are designed to explore the impact of violating assumption A3 on model performance. Note that for all the scenarios mentioned above, the LGM does not include nonlinear or interaction terms when predicting response propensity scores. Therefore, WT-LGM may suffer from the issue of model misspecification. We specify the number of trees $B=100$, posterior draws $D=1,000$ and burn-in draws $100$. We use the default prior specifications $\nu = 3$, $\alpha = 0.95$, $\beta=2$, $k = 2$ when building BART and rBART models in the simulation and real application.

\subsection{\large Simulation results}
\vspace{0.05in}

\begin{figure}
\centerline{\includegraphics[width=\linewidth]{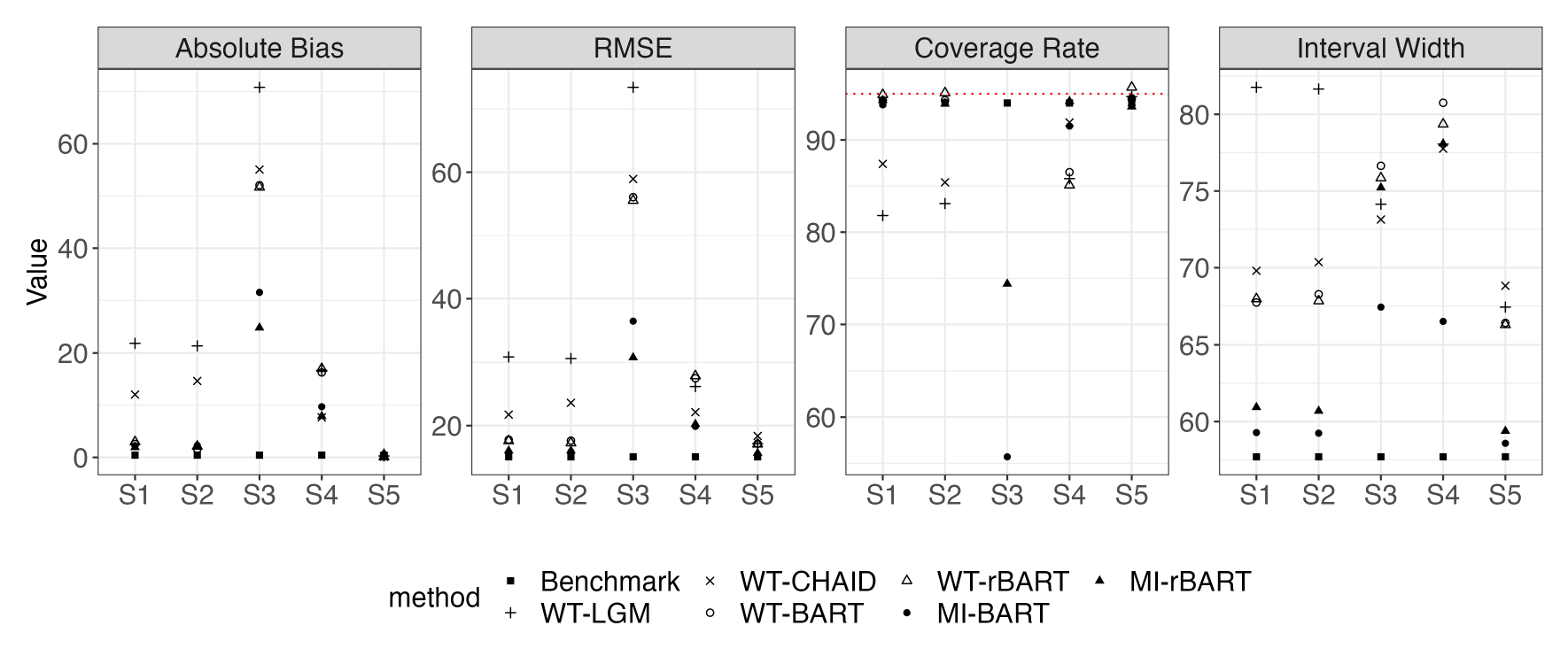}}
\caption{Simulation results with $D=10$ imputations across five simulation scenarios. WT-LGM, WT-CHAID, WT-BART, and WT-rBART denote weighting methods based on the Phase II sample, while MI-BART and MI-rBART represent tree-based imputation methods. Experimental conditions are: S1 - low-dimensional setting with all assumptions satisfied; S2 - high-dimensional setting with all assumptions satisfied; S3 - high-dimensional setting with violation of assumption (A3) for an outcome-related variable; S4 - high-dimensional setting with violation of assumption (A3) for a non-outcome-related variable; and S5 - high-dimensional setting with no overlap between variables related to $y$ and $\pi$. Coverage rates for some weighted estimators with low values are omitted.}
\label{visual-simulation}
\end{figure}

Figure~\ref{visual-simulation} shows the simulation results with $D=10$ imputations, where the y-axis represents the value of each metric. The benchmark estimator, which assumes that the Phase II sample is equivalent to the full Phase I sample, has the lowest absolute bias and RMSE, the narrowest interval width, and coverage rates closest to the nominal level across all scenarios. This result is expected, as this approach does not require imputing the Phase I sample and thus avoids imputation error. 

The MI-based estimators exhibit similar bias and RMSE to the benchmark estimator in scenarios S1, S2, and S5 but show increased bias and RMSE in scenarios S3 and S4.  
Overall, the MI-based estimators consistently outperform the weighting estimators across all scenarios in terms of bias, RMSE, and coverage. The 95\% CI coverage rates for the two MI-based estimators are close to the nominal level across all scenarios except S3, where all the weighting methods have poor coverage rates and their results are therefore omitted from Figure~\ref{visual-simulation}. 

The two MI-based estimators, using BART and rBART imputation models, yield similar bias and RMSE across all scenarios except S3, where MI-rBART shows lower bias and RMSE and achieves coverage rate closer to the nominal level than MI-BART. Furthermore, MI-rBART consistently produces wider interval widths than MI-BART in all scenarios. When the number of imputations is increased from $10$ to $1000$, the simulation results remain largely unchanged, although the interval widths of the MI-based estimators decrease slightly with more imputations. Results for $D=20, 50, 1000$ are presented in eFigures 1-3 in the supplementary materials. 

The weighting methods using BART and rBART outperform WT-LGM and WT-CHAID, exhibiting much lower bias and RMSE, narrower 95\% CIs, and coverage rates closer to the nominal level. This improvement is expected, as BART and rBART can handle high-dimensional covariates and can effectively capture nonlinear relationships and interaction effects in the response propensity model without requiring pre-selection of ``important'' predictors. In scenarios S1, S2, and S5, WT-BART and WT-rBART show similar bias but slightly higher RMSE compared to their MI-based counterparts. Although their 95\% CIs achieve coverage rates near the nominal level, the intervals are much wider than those from the MI-based estimators. In contrast, under scenarios S3 and S4, both WT-BART and WT-rBART yield substantially higher bias and RMSE and poor coverage rates relative to the MI-based estimators.

When the assumption A3 is violated by the sparse data in the tails of outcome-related variable $x_2$ under scenario S3, both the imputation and weighting methods perform worse than the other scenarios, although the MI-based estimators still perform better than the weighting estimators with a much smaller bias and RMSE. For scenario S4 with assumption A3 violated due to the sparse data in the tails of $x_3$, a variable not correlated with $y$, the weighting and MI estimators using BART and rBART still perform well, with moderately increased bias and RMSE but much wider 95\% CIs than in scenarios S1 and S2.

\begin{figure}
\centerline{\includegraphics[width=\linewidth]{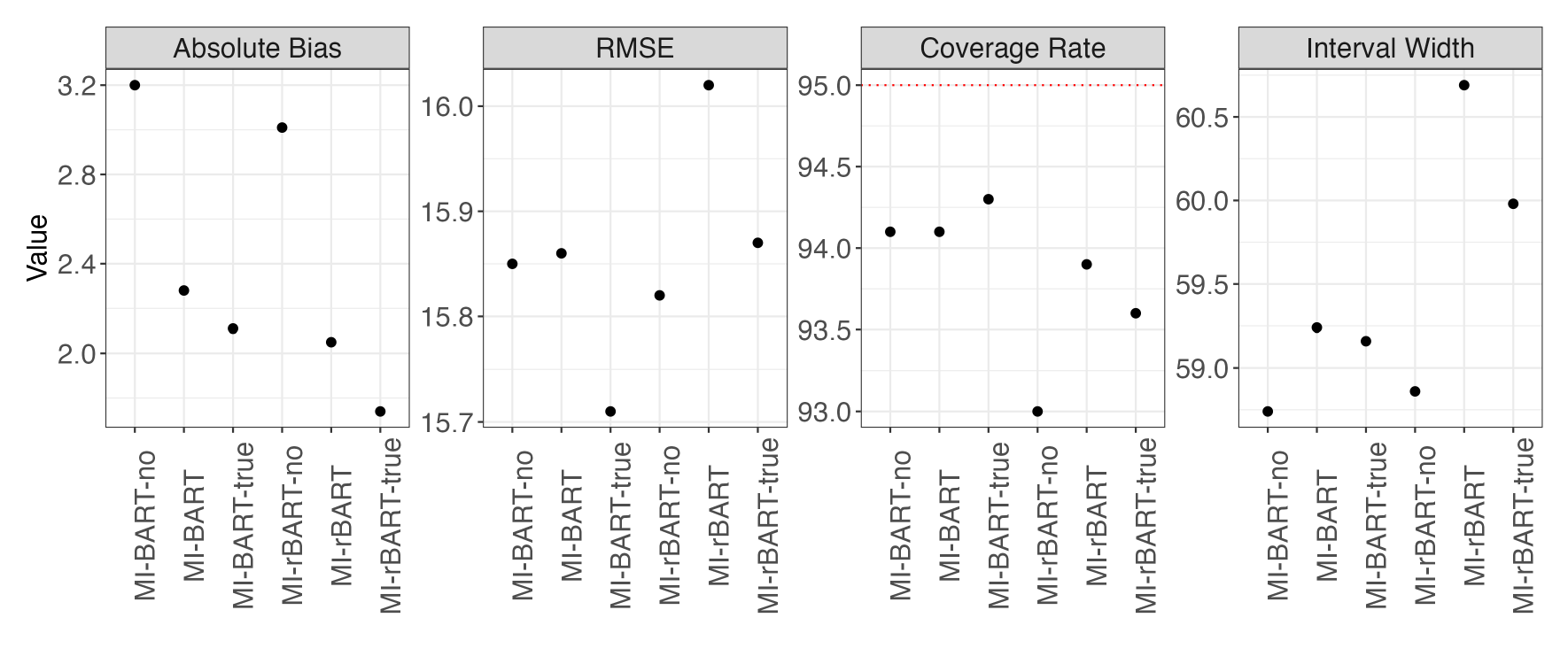}}
\caption{Simulation results evaluating the impact of including estimated subsample weighting adjustments in the imputation model ($D=10$, Scenario S2). MI-(r)BART: includes $a_i$ calculated from estimated response propensity; MI-(r)BART-no: excludes $a_i$ from the imputation model; MI-(r)BART-true: includes $a_i$ calculated from true response propensities. }
\label{propensity}
\end{figure}

Figure~\ref{propensity} presents the results of the sensitivity analysis under Scenario S2, comparing MI-(r)BART with two variations: one that removes the subsample selection and nonresponse weighting adjustments $a_i$ from the imputation models, denoted MI-(r)BART-no, and another that calculates $a_i$ using the inverse of the true response propensities, denoted MI-(r)BART-true, under simulation scenario S2. Omitting $a_i$ from the imputation models leads to larger biases in both MI-BART and MI-rBART estimates, and a slightly lower coverage rate for MI-rBART. In contrast, calculating $a_i$ based on the true response propensities results in estimates with slightly smaller bias and RMSE, as well as narrower 95\% confidence intervals.

\section{\large A subcohort cellphone survey in Uganda}
\vspace{0.1in}

We apply the proposed tree-based imputation method to estimate COVID-19 vaccination coverage among adults aged 18 and older in Uganda, using the subcohort cellphone survey sampled from the larger UPHIA2020 sample. A total of 124 covariates from the UPHIA survey were included in the analysis, comprising 37 house-level variables and 87 individual-level variables. The household-level variables included household location (region, urban/rural), structural characteristics (floor, roof, walls, number of rooms), assets (livestock, household items, wealth quintile), economy indicators (income sources), water and sanitation facilities (toilet type, water source), and the number of children in the household. Individual-level covariates covered demographics (age, sex, education, ethnic group, religion), socioeconomic indicators (employment, marital status), health status, and variables related to HIV prevention and risk behaviors. 

Unweighted and weighted descriptive statistics for a selected subset of the covariates are summarized in Table \ref{tab1}. We compared the distributions of these covariates between the participants in UPHIA2020 aged over 18 (N=23,214) and the cellphone survey subcohort, considering both primary respondents only (n = 831) and the combined primary and secondary respondents (n=887). Compared to the UPHIA2020 sample, primary respondents in the cellphone survey were more likely to reside in urban areas (39.4\%  vs. 31.3\%), be male (50.5\% vs. 41.0\%), be married (70.2\% vs. 62.6\%), have worked in the past 12 months (64.0\% vs. 57.6\%); were less likely to be 24 years or younger (14.6\% vs. 26.9\%), have no formal education (7.1\% vs. 14.6\%), and fall into the lowest wealth quantile (15.9\% vs. 29.4\%). 

The distributions remain similar with and without the inclusion of secondary respondents, though including the secondary respondents makes the cellphone subcohort covariates resemble the UPHIA2020 sample more closely. These discrepancies highlight the potential for bias in estimating vaccination coverage using the cellphone subcohort if proper adjustments are not applied. One option to adjust this is to construct weights for the cellphone sample. These were constructed by multiplying the final person-level weights from UPHIA2020 by subcohort adjustment factors, which were derived using the inverse of estimated propensity scores obtained from a probit BART model. When applied, the weighted distributions of these covariates in the cellphone sample align more closely with those in the UPHIA2020 sample. Notably there was still considerable deviation in the weighted estimates for the subcohort and the UPHIA sample when estimating the proportion who were aged 18-24 (26.2\% vs. 30.3\%) and those who fall into the lowest wealth quantile (16.6\% vs. 20.8\%). 

\begin{table}[t]
\centering
\caption{Unweighted and weighted descriptive statistics for selected covariates from the UPHIA2020 survey: comparison between the full UPHIA2020 survey and the cellphone subcohort using primary and combined respondents. Values in parentheses represent unweighted and weighted percentages, respectively.}

\resizebox{0.9\columnwidth}{!}{
\begin{tabular}{lrrr}
\toprule
\textbf{Characteristics} & \textbf{Primary sample} & \textbf{Primary + Secondary sample} & \textbf{UPHIA sample} \\

\textbf{} & (N = 831) & (N = 887) & (N = 23,214) \\
\bottomrule
Location &  & \\
\hspace{1em}Urban & 327 (39.4\%, 34.9\%) & 332 (37.4\%, 34.0\%) & 7,260 (31.3\%, 33.6\%) \\
\hspace{1em}Rural & 504 (60.6\%, 65.1\%) & 555 (62.6\%, 66.0\%) & 15,954 (68.7\%, 66.4\%) \\
Gender &  & \\
\hspace{1em}Male & 421 (50.5\%, 48.9\%) & 430 (48.5\%, 47.5\%) & 9,515 (41.0\%, 47.0\%) \\
\hspace{1em}Female & 410 (49.3\%, 51.1\%) & 457 (51.5\%, 52.5\%) & 13,699 (59.0\%, 53.0\%) \\
Age &  & \\
\hspace{1em}18-24 & 121 (14.6\%, 26.2\%) & 137 (15.4\%, 25.7\%) & 6,240 (26.9\%, 30.3\%) \\
\hspace{1em}25-34 & 278 (33.5\%, 31.9\%) & 294 (33.1\%, 33.3\%) & 6,265 (27.0\%, 29.3\%) \\
\hspace{1em}35-44 & 193 (23.2\%, 17.1\%) & 203 (22.9\%, 17.5\%) & 4,399 (18.9\%, 17.7\%) \\
\hspace{1em}45-54 & 131 (15.8\%, 12.6\%) & 134 (15.1\%, 11.3\%) & 2,971 (12.8\%, 11.0\%) \\
\hspace{1em}55-64 & 64 (7.7\%, 6.8\%) & 68 (7.7\%, 7.0\%) & 1,711 (7.4\%, 6.4\%) \\
\hspace{1em}65+ & 44 (5.3\%, 5.4\%) & 51 (5.7\%, 5.3\%) & 1,628 (7.0\%, 5.3\%) \\
Education & & \\
\hspace{1em}No education & 59 (7.1\%, 8.7\%) & 70 (7.9\%, 8.9\%) & 3,382 (14.6\%, 9.5\%) \\
\hspace{1em}Primary & 400 (48.1\%, 49.9\%) & 428 (42.3\%, 49.4\%) & 11,803 (50.8\%, 50.9\%) \\
\hspace{1em}Secondary & 252 (30.3\%, 30.3\%) & 267 (30.1\%, 30.8\%) & 5,940 (25.6\%, 29.6\%) \\
\hspace{1em}More than secondary & 120 (14.4\%, 11.1\%) & 121 (13.6\%, 10.7\%) & 2,070 (8.9\%, 9.9\%) \\
\hspace{1em}Missing & 0 (0.0\%, 0.0\%) & 1 (0.1\%, 0.1\%) & 19 (0.1\%, 0.1\%) \\
Worked in past 12 months & & \\
\hspace{1em}Yes & 532 (64.0\%, 63.2\%) & 562 (63.4\%, 62.2\%) & 13,378 (57.6\%, 61.9\%) \\
\hspace{1em}No & 299 (36.0\%, 36.8\%) & 325 (36.6\%, 37.8\%) & 9.827 (42.3\%, 38.1\%) \\
\hspace{1em}Missing & 0 (0.0\%, 0.0\%) & 0 (0.0\%, 0.0\%) & 9 (0.0\%, 0.0\%) \\
Marriage &  & \\
\hspace{1em}Never Married & 107 (12.9\%, 21.5\%) & 114 (12.9\%, 22.0\%) & 4,132 (17.8\%, 21.3\%) \\
\hspace{1em}Married/living together & 583 (70.2\%, 61.3\%) & 626 (70.6\%, 61.1\%) & 14,522 (62.6\%, 60.6\%) \\
\hspace{1em}Divorced/Separated & 110 (13.2\%, 13.3\%) & 115 (13.0\%, 12.7\%) & 2,963 (12.8\%, 12.8\%) \\
\hspace{1em}Widowed & 31 (3.7\%, 3.9\%) & 32 (3.6\%, 4.2\%) & 1,569 (6.8\%, 5.1\%) \\
\hspace{1em}Missing & 0 (0.0\%, 0.0\%) & 0 (0.0\%, 0.0\%) & 28 (0.1\%, 0.1\%) \\
Wealth quantile & & \\
\hspace{1em}Lowest & 132 (15.9\%, 16.6\%) & 150 (16.9\%, 17.8\%) & 6,835 (29.4\%, 20.8\%) \\
\hspace{1em}Second & 148 (17.8\%, 22.5\%) & 164 (18.5\%, 22.2\%) & 4,445 (19.1\%, 20.3\%) \\
\hspace{1em}Middle & 156 (18.8\%, 21.4\%) & 173 (19.5\%, 21.5\%) & 4,383 (18.9\%, 21.4\%) \\
\hspace{1em}Fourth & 185 (22.3\%, 18.2\%) & 187 (21.1\%, 18.3\%) & 3,710 (16.0\%, 18.6\%) \\
\hspace{1em}Highest & 210 (25.3\%, 21.2\%) & 213 (24.0\%, 20.2\%) & 3,834 (16.5\%, 18.8\%) \\
\hspace{1em}Missing & 0 (0.0\%, 0.0\%) & 0 (0.0\%, 0.0\%) & 7 (0.0\%, 0.0\%) \\
\bottomrule
\end{tabular}}
\label{tab1}
\end{table}

Missing data in the covariates was a minor concern in this application. To address it, we applied multiple imputation by chained equations using the ``mice'' package in R to impute missing values in UPHIA2020. Five imputed data sets were generated for analysis. 

To estimate COVID-19 vaccination coverage in the population, we fit both BART and rBART models using data from the cellphone survey, considering both primary respondents only and the combined primary and secondary respondents. The binary indicator of COVID-19 vaccination served as the outcome. Each BART model included all covariates from UPHIA2020, as well as survey design variables: strata, clusters, and the final person-level sample weight from the UPHIA2020 survey. After model fitting, we used posterior draws to predict the vaccination status of UPHIA2020 participants who were not included in the cellphone survey. 

The estimated vaccination coverage for the population was then obtained by calculating the weighted mean of the vaccination variable, using the observed values for cellphone respondents and the predicted values for non-respondents, with weights derived from the UPHIA2020 survey. From each of the five imputed UPHIA2020 data sets, we generated 50 estimates of vaccination coverage, yielding a total of 250 estimates, which were averaged to produce the MI-BART estimate. Inference was based on 95\% confidence intervals, computed using Rubin’s rules for variance estimation. We repeated the analysis using both the combined primary and secondary respondents and the random-intercept BART model (rBART). For comparison, we also calculated weighted estimates of vaccination coverage using the subcohort weights, with propensity scores estimated from both BART and rBART models.

\begin{figure}[ht]
\centerline{\includegraphics[width=1\linewidth]{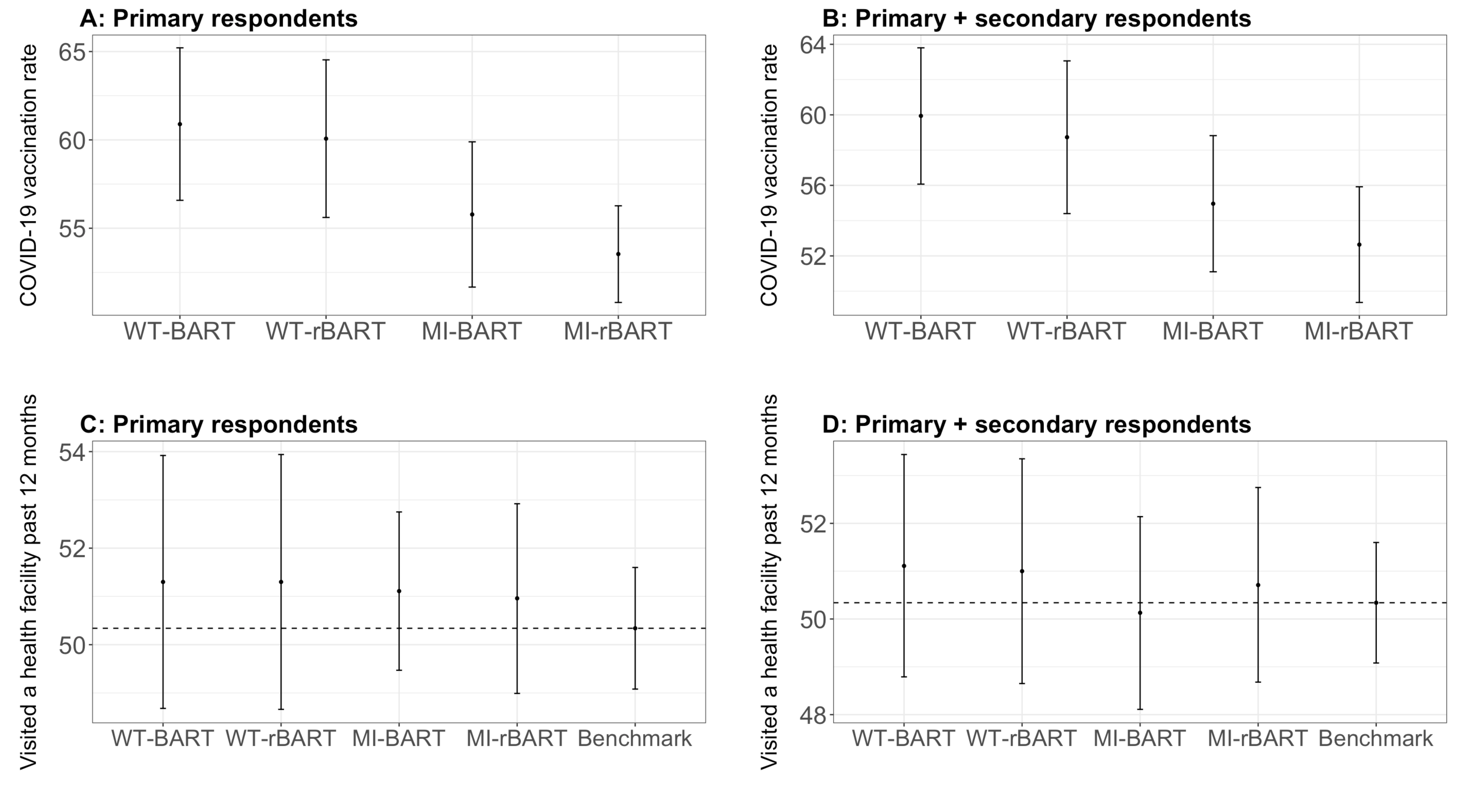}}
\caption{Results for the Uganda subcohort cellphone survey application. Comparison of MI-BART and MI-rBART with WT-BART and WT-rBART in estimating two outcomes: COVID-19 vaccination coverage and the percentage of individuals who visited a health facility in the past 12 months. Analyses were conducted with and without including secondary respondents. Results are shown as point estimates with 95\% confidence intervals. Benchmark estimates from the full UPHIA2020 sample are included for the second outcome as a reference.}
\label{fig:subsample}
\end{figure}

The results for estimating COVID-19 vaccination coverage are presented in Figure~\ref{fig:subsample} and the Supplementary eTable 1.  According to the U.S. Centers for Disease Control and Prevention, by the summer of 2022, approximately 57\% of Ugandans aged 18 and older had received their primary COVID-19 vaccinations \citep{cdc2022uganda}. Since the cellphone survey was conducted in March 2022, the actual coverage at that time was likely lower. Using only primary respondents in the cellphone survey subcohort, we estimated that 53.5\% (95\% CI: 50.8\%, 56.3\%) of adults in Uganda were fully vaccinated against COVID-19 based on the MI-rBART model, and 55.8\% (95\% CI: 51.7\%, 59.9\%) using MI-BART. The corresponding weighted estimates were 60.1\% (95\% CI: 50.6, 65.2\%) using WT-rBART and 60.9\% (95\% CI: 50.7\%, 65.2\%) using WT-BART. Both weighted estimates exceed the CDC's reported value for summer 2022, suggesting potential upward bias in the weighting-based approaches. Including secondary respondents in the analysis yields slightly lower estimates of vaccination coverage, bringing the weighted results slightly closer to the CDC figure. This suggests that including secondary respondents, who did not own a cellphone, may help reduce bias, though the impact appears modest.  

In both the UPHIA2020 and the cellphone survey, participants were asked, “Have you seen a health worker in a health facility in the last 12 months?” (yes or no). We used this variable to conduct a validation study assessing the performance of the BART models used in this analysis, assuming that the underlying population hasn't changed from the initial survey point to the March 2022 cellphone survey. To validate the results, we repeated vaccination coverage estimation procedure, replacing the outcome to ``visited a health facility in past 12 months''. The resulting estimates were compared to a benchmark, derived from the observed data in the full UPHIA2020 parent cohort.  

The results of this validation study are presented in Figure~\ref{fig:subsample} and the Supplementary eTable 2. Using the full UPHIA2020 sample, we estimated that 50.3\% (95\% CI: 49.1, 51.6) of adults had visited a health facility in the last 12 months. All four subcohort-based methods produced similar, though slightly higher estimates, yielding the closest match to the benchmark with MI-rBART using only primary respondents and with MI-BART using the combined sample. Specifically,  MI-rBART estimated 51.0\% (95\% CI: 49.0\%, 52.9\%) of adults had visited a health facility in the past 12 months using only primary respondents, and MI-BART estimated 50.1\% (95\% CI: 48.1\%, 52.1\%) when including both primary and secondary respondents. Both MI-BART and MI-rBART produced narrower 95\% confidence intervals than the corresponding weighted estimators. Consistent with the findings for vaccination coverage, including secondary respondents led to slightly lower estimates, bringing the results closer to the benchmark.

We also examined the importance of each covariate in predicting COVID-19 vaccination status by analyzing the frequency with which each variable appeared in the trees of the BART model. Notably, both region of residence and cluster indicators ranked among the top 15 most frequently occurring predictors. This finding highlights the importance of survey design variables in predicting COVID-19 vaccination status and provide valuable insights into their role within the context of our study.  

\section{\large Discussion}
\vspace{0.1in}

The primary objective of this paper is to improve statistical inference for population quantities in two-phase complex survey designs by leveraging the rich auxiliary information collected in the Phase I sample. Weighted analyses are the most commonly used approach to estimate population quantities in two-phase designs, adjusting for the nonresponse and selection bias. However, these methods have several limitations. First, high-dimensional auxiliary variables are often present in the Phase I sample, posing challenges in constructing subsample weights for the Phase II sample based on all available auxiliary variables collected in Phase I. The resulting weighted estimator may be biased if the subsample weights are not properly specified. Second, weighted analyses based on subsample weights typically fail to account for the relationships between auxiliary variables and outcomes, thereby overlooking useful predictive information. Finally, some respondents may have very small estimated response propensities, leading to extremely large weights and, consequently, high variance in the weighted estimators. This issue is especially pronounced in two-phase designs, where the final weights are calculated as the product of the Phase I weights and subsample adjustment factors.  

In contrast, by treating the outcomes for units in the Phase I sample but not in the Phase II sample as missing, we can use imputation to fill in these unobserved outcomes.
Imputation-based methods can improve the efficiency of survey estimates from the Phase II sample, particularly when the Phase I sample includes important predictors of the outcomes of interest. However, the quality of imputation depends on the imputation model; a misspecified model can lead to poor results. Machine learning-based imputation models are appealing in this context, as they can handle high-dimensional auxiliary variables and are more robust to model misspecification \citep{chipman2007bayesian}. In this paper, we develop the MI-BART and MI-rBART methods for survey inference in two-phase complex survey designs. These methods involve first using a BART or rBART model to multiply impute outcomes that are only observed in the Phase II sample, followed by a multiple imputation analysis using the completed Phase I data sets. 

Simulation results show that the proposed MI-BART and MI-rBART methods outperform traditional weighted analyses based on the subsample weights when estimating population means from the Phase II sample in a two-phase design. Specifically, they achieve lower bias, reduced RMSE, narrower confidence intervals, and coverage rates closer to nominal level. MI-BART and MI-rBART perform comparably in most settings, while in general MI-BART yields slightly narrower 95\% confidence intervals compared to MI-rBART. The simulations also highlight the advantage of tree-based models in subsample weighted analyses. Both WT-BART and WT-rBART exhibit smaller bias and RMSE compared to WT-LGM and WT-CHAID, with WT-LGM showing the poorest due to model misspecification.

Building on a well-designed and executed national HIV survey, we conducted a subcohort cellphone survey to estimate national COVID-19 vaccination coverage in Uganda. The results demonstrate that our proposed MI-BART and MI-rBART methods effectively reduce bias in estimates derived from the cellphone survey subcohort when compared to traditional weighting approaches, by leveraging the rich auxiliary information collected in the parent survey. In addition, our validation study further confirmed the advantages of these methods, as they produced estimates closer to the benchmark derived from the full parent cohort data. Overall, this study suggests that a two-phase survey design, combined with our proposed analysis approach, offers a cost-effective alternative for population inference—particularly in settings where designing and implementing a new probability survey is either too costly or infeasible. 

Although BART, like many machine learning methods, is often viewed as a ``black-box'' due to limited interpretability, our empirical results demonstrate that it can still yield valuable insights. By examining variable importance measures, key drivers of the outcome can be identified, providing practical guidance for policy researchers aiming to design targeted health interventions. Moreover, when the Phase I sample is obtained through multistage probability sampling, it is essential that both MI-BART and MI-rBART incorporate survey design features, such as strata, clusters, Phase I sample weights, and subsample weight adjustments, as covariates in the imputation models to ensure valid inference \citep{chen2015}. 

A key assumption of the MI-based methods is that the Phase II sampling mechanism is ignorable, given the auxiliary information and design variables. This assumption becomes more plausible when a large number of auxiliary variables are available, as is often the case in two-phase designs. For valid imputation, the ranges of auxiliary variables, particularly those predictive of the outcome,  should be similar between the Phase I and Phase II samples. If the Phase II sample has a narrower range of key predictors, the MI-based methods may produce biased estimates, though they still outperforming weighting methods, as demonstrated in Scenario S3 of our simulation study.

\citet{liu2023inference} showed that including the estimated propensity of inclusion as a covariate in the tree-based imputation model can help guard against model misspecification in BART, especially when key predictors of $Y$ are omitted. Our simulations support this finding: excluding subsample weighting adjustments from the imputation models leads to slightly larger biases in both MI-BART and MI-rBART estimates even though all the important predictors are included. Since the true subsample propensity of inclusion is typically unknown, it is common practice to include the estimated propensity in the imputation model.  Our results indicate that using the estimated rather than the true propensity scores leads to only a slight increase in bias and RMSE.  Moreover, because the estimated propensity is used solely as a candidate predictor in the outcome model, ignoring its uncertainty has limited impact on the estimation of population quantities.

In estimating population mean using the Phase II sample under two-phase complex survey designs, we adopt MI inference with Rubin's variance estimation formula. Specifically, we first estimate the variance of each imputed data set using the Taylor series linearization to account for the effects of complex survey designs of the Phase I sample. We then incorporate imputation uncertainty using Rubin's MI formula for variance estimation. This approach differs from that of \citet{liu2023inference}, who analyzed simpler, one-phase samples obtained via single-stage sampling from a finite population. Our simulations show that direct inference based on the posterior distribution produces overly narrow 95\% probability intervals with coverage rates below the nominal level (\emph{see} Supplementary eFigure 4).

The tree-based MI methods for two-phase designs have applications beyond survey settings and can be extended to clinical trials and epidemiologic studies to generalize findings from a sample to a broader target population. In this paper, we focus on BART models due to their strong predictive performance and ability to support Bayesian posterior inference. However, other Bayesian machine learning methods with robust predictive capabilities may also be appropriate for this framework. 

Under Scenario S3, all models exhibited degraded performance, although the two tree-based MI methods outperformed the alternatives. This reflects a classic covariate shift problem in machine learning, where the distribution of key predictors differs between the training and prediction samples. A promising future direction is the development of a hybrid BART model that incorporates a parametric component for covariates subject to covariate shift, while modeling the remaining variables nonparametrically within the BART framework.

\section{Data and Code Availability Statement}

\sloppy R Code for the simulations are available at \url{https://github.com/cherylwaal/MI-BART-in-two-phase-designs}. The UPHIA2020 dataset can be accessed via \url{https://phia-data.icap.columbia.edu/datasets?country_id=7}.

\section{Competing interests}
No competing interest is declared.

\section{Acknowledgments and Funding} 
We thank the Editor, Associate Editor, and the referees for their thoughtful and constructive comments, which have significantly improved the quality of this paper. This work was partially supported by the National Institutes of Health (R01AG067149, R01ES035784). Wang is supported by the PhD scholarship from the Duke-NUS Medical School, Singapore.

\end{spacing}
\end{document}